\documentclass[aps,prl,twocolumn,superscriptaddress,showpacs]{revtex4}

\usepackage{graphicx}
\usepackage{dcolumn}
\renewcommand{\vec}[1]{\mbox{\boldmath $#1$}}
\usepackage{color}

\begin{document}
\preprint{\#1}

\title{Charge-Order Pattern of the Low-Temperature Phase of NaV$_{2}$O$_{5}$ Uniquely Determined by Resonant X-Ray Scattering from Monoclinic Single Domain}

\author{Kenji Ohwada}
\affiliation{Synchrotron Radiation Research Center (SPring-8), Japan Atomic Energy Research Institute, Kohto, Hyogo 679-5148, Japan}

\author{Yasuhiko Fujii}
\altaffiliation[Present address: ]{Neutron Science Research Center, Japan Atomic Energy Research Institute, Tokai, Ibaraki 319-1195, Japan}
\author{Yuya Katsuki}
 \altaffiliation[Present address: ]{Accenture Co., Ltd., Minato-ku, Tokyo 107-8672, Japan}
 \author{Jiro Muraoka}
 \altaffiliation[Present address: ]{All Nippon Airways Co., Ltd., Minato-ku, Tokyo 105-7133, Japan}
\affiliation{Neutron Science Laboratory, Institute for Solid State Physics, The University of Tokyo, 106-1, Shirakata, Tokai, Ibaraki 319-1106, Japan}

\author{Hironori Nakao}
\author{Youichi Murakami}
\affiliation{Department of Physics, Graduate School of Science, Tohoku University, Sendai, Miyagi 980-8578, Japan}

\author{Hiroshi Sawa}
\author{Emi Ninomiya}
\affiliation{Photon Factory, Institute of Materials Structure Science, KEK, Tsukuba, Ibaraki 305-0801, Japan}
\altaffiliation[Present address: ]{TDK Co., Ltd., Chuo-ku, Tokyo 103-8272, Japan}

\author{Masahiko Isobe}
\author{Yutaka Ueda}
\affiliation{
Materials Design and Characterization Laboratories, Institute for Solid State Physics, The University of Tokyo, Kashiwa, Chiba 277-8581, Japan}

\date{\today}

\begin{abstract}
The present resonant x-ray scattering from each of monoclinically-split single domains of NaV$_{2}$O$_{5}$ has critically enhanced contrast between V$^{4+}$ and V$^{5+}$ ions strong enough to lead to unambiguous conclusion of the charge-order pattern of its low-temperature phase below $T_{\rm c}$ = 35 K.  The zig-zag type charge-order patterns in the $ab$-plane previously confirmed have four kinds of configurations (A, A', B and B') and the stacking sequence along the $c$-axis is determined as the AAA'A' type by comparison with model calculations.  By assigning the A and A' configurations to Ising spins, one can reasonably understand the previously discovered "devil's staircase"-type behavior with respect to the modulation of the layer-stacking sequences at high pressures and low temperatures, which very well resembles the global phase diagram theoretically predicted by the ANNNI model.
\end{abstract}

\pacs{71.27.+a, 61.10.-i}
\maketitle

\begin{figure}[t]
\begin{center}
\includegraphics[width=0.95\linewidth]{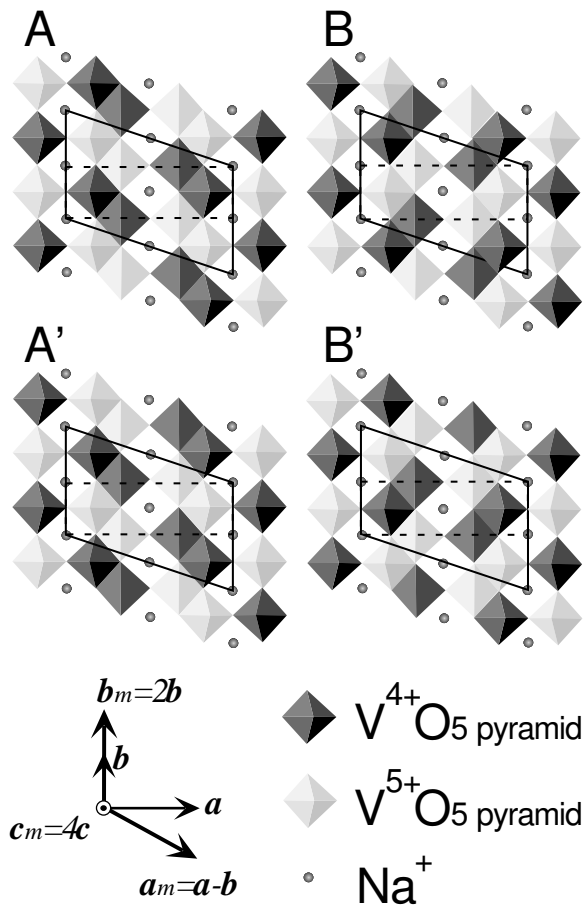}
\end{center}
\caption{The zig-zag type charge-order pattern with respect to configuration of the V$^{4+}$O$_{5}$ (black) and V$^{5+}$O$_{5}$ (white) pyramids laying in the ab-plane previously confirmed~\cite{sawa,ninomiya}. There are four kinds of possible in-plane configurations as denoted as A, A', B and B'.  Dotted and solid lines represent the crystallographic unit cells, orthorhombic ($a \times b \times c$) above $T_{\rm c}$ = 35 K and monoclinic ($a_{\rm m} \times b_{\rm m} \times c_{\rm m}$) below Tc.
}
\label{fig:1}
\end{figure}

 Since the discovery of the spin-Peierls-like phase transition of NaV$_{2}$O$_{5}$ at $T_{\rm c}$ = 35 K~\cite{isobe1}, its low-temperature structure has been a controversial question.
 NaV$_{2}$O$_{5}$ is well described by a system of quarter-filled two-leg spin ladders~\cite{new-xray1,seo-fukuyama,tsuda}, running along the $b$-axis of its orthorhombic structure above $T_{\rm c}$ ($a = 11.3$, $b = 3.65$, $c = 4.8$ $\rm{\AA}$). All vanadium ions have a nominal valence state of +4.5 (V$^{4.5+}$) at room temperature; one electron is distributed on one V-O-V rung parallel to the $a$-axis. At $T_{\rm c}$ = 35 K, NaV$_{2}$O$_{5}$ undergoes a novel cooperative phase transition associated with its charge disproportionation as 2V$^{4.5+}$$\rightarrow$V$^{4+}$(spin state S = 1/2)+V$^{5+}$(S = 0)~\cite{ohama1}, lattice dimerization as indexed by a $2a \times 2b \times 4c$ supercell~\cite{fujii1} and spin-gap formation ($\Delta$ = 9.8 meV)~\cite{fujii1,yosihama}.
 The primal analysis of its low-temperature structure based on the space group C$^{18}_{2v}$-Fmm2~\cite{ludecke,boer} suggested three different electronic states of the V sites, the charge-ordered V$^{4+}$, V$^{5+}$ and disordered V$^{4.5+}$. However, such a charge distribution is incompatible with experimental results by $^{51}$V NMR~\cite{ohama1} and resonant x-ray scattering (RXS) measurements~\cite{nakao}. Furthermore, $^{23}$Na NMR spectral measurements~\cite{ohama2} showed eight independent Na sites, in contrast to only six Na sites led by the space group Fmm2. Thus, the low temperature (LT) structure and its related charge distribution (charge-order pattern) of NaV$_{2}$O$_{5}$ have long been a hot issue.
 In 2002, Sawa {\it et~al.} succeeded in observation of Bragg peak splitting below $T_{\rm c}$, leading to the fact that the LT phase is {\it monoclinic}~\cite{sawa,ninomiya}. It was then determined by them that the LT monoclinic unit cell is constructed as $\rm{(}\it{a}-\it{b}\rm{)} \times \rm{2} \it{b} \times \rm{4} \it{c}$ with the space group C$^{3}_{2}$-A112 as shown in Fig.~\ref{fig:1}. By taking into account two kinds of monoclinically-split domains, they obtained a completely different structure from the previously conjectured one with the Fmm2. The bond valence sum method applied to the new structure results in that the V sites are clearly categorized into two groups as the charge-ordered V$^{4+}$ and V$^{5+}$ with a zig-zag pattern in a ladder as seen in Fig.~\ref{fig:1}, where a VO$_{5}$ pyramid containing V$^{4+}$ is colored dark while one for V$^{5+}$ bright. The most left two VO$_{5}$ pyramid linkage shared with each corner forms a ladder running along the $b$-axis. Thus obtained LT structure is consistent with the previous resonant x-ray~\cite{nakao} and NMR~\cite{ohama1,ohama2} data, in striking contrast to the structure previously reported, which included the disordered V$^{4.5+}$ sites~\cite{ludecke,boer}. 
%
 Figure~\ref{fig:1} also shows four kinds of charge-order patterns in the $ab$ plane (called A, A', B and B') possibly obtained from the charge-ordering process upon phase transition in which one electron equally shared by two V ions in a rung above $T_{\rm c}$ is localized at either of two V ions below $T_{\rm c}$. In the A pattern, for example, the V$^{4+}$ is located at each rung within a ladder in a zig-zag manner. All ladders have such a zig-zag charge-order but the relationship between two adjacent ladders causes four kinds of patterns in a single layer as displayed. 
 Since the layer-stacking direction along the $c$-axis becomes quadruple below $T_{\rm c}$, the sequence of these four kinds of layers is a central issue to be solved for explaining all of the observed physical properties, consistently.
 Recently Sawa {\it et~al.}~\cite{sawa} carried out an x-ray diffraction experiment to collect intensity data from a mixed domain sample and to analyze them based on the monoclinic symmetry. Thus they reported that the ABA'B' stacking sequence gave the best fit to the observation. However, there is another candidate for a possible stacking sequence as AAA'A' which also satisfies the space group A112. It should be noted that in the domain-averaged intensity-data analysis, the intensity for a combination of the sequences ABA'B'+BAB'A' with an equal domain distribution is found to be the same as that of AAA'A'+BBB'B'. This fact means that a unique structure cannot be obtained unless intensity data is collected from a monoclinic single domain. 
Such a problem was also pointed out by Grenier {\it et al.} who made the RXS study with domain averaged data~\cite{grenier}.
 In spite of such intensive experimental studies for last several years, however, the LT structure and charge-order pattern of NaV$_{2}$O$_{5}$ have not been determined satisfactorily. Once a single domain is available and the charge contrast between V$^{4+}$ and V$^{5+}$ is observable enough near the absorption edge of V ion, it should be easy to distinguish between the stacking sequences ABA'B' and AAA'A'. In this paper, we have reported unambiguously determined charge-order pattern based on our RXS experiments by using each of two kinds of monoclinically-split single domains.

 The RXS measurements were performed by synchrotron x-rays at beam lines BL-4C and 9C at Photon Factory of KEK. Incident x-rays were monochromatized with a Si(111) double crystal monochromator. X-ray energy was varied across the V K-absorption-edge (5.47 keV) which was calibrated with the absorption edge of a V metal foil. A very small single crystal of NaV$_{2}$O$_{5}$ with a size of 58 $\times$ 92 $\times$ 36 $\rm{{\mu}m}^{3}$ ($a$$\times$$b$$\times$$c$) grown by the same method as previously reported~\cite{isobe1, isobe2} was mounted on a diamond sample holder~\cite{diamond} by using a very small amount of silicone grease so as not to apply any physical stress. Its $c$-axis was set perpendicular to the diamond surface 


\begin{figure}[t]
\begin{center}
\includegraphics[width=0.95\linewidth]{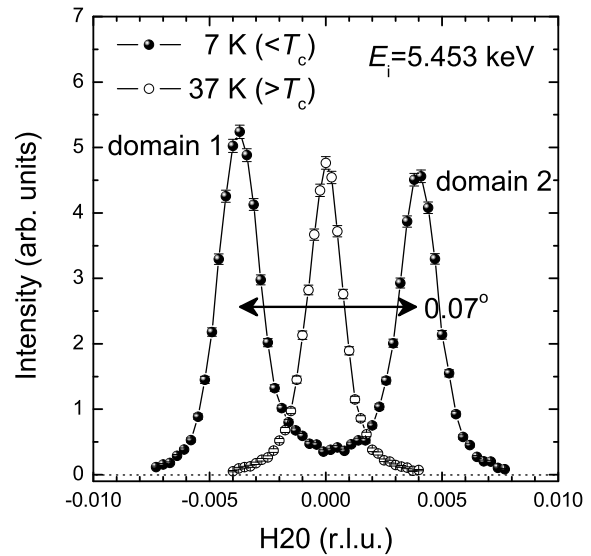}
\end{center}
\caption{Observed peak-splitting of the 020 fundamental Bragg reflection directly evidences the orthorhombic-to-monoclinic phase transition at $T_{\rm c}$ = 35 K. The splitting angle 0.07$^{\circ}$ is consistent with the previous reports~\cite{sawa,ninomiya}. The integrated intensity ratio of the monoclinically-split domains 1 and 2 gives their volume ratio of 55:45.}
\label{fig:2}
\end{figure}
Figure~\ref{fig:2} shows the peak profiles of the 020 fundamental Bragg reflection taken with $E_{\rm i} = 5.453$ keV below and above $T_{\rm c}$, 7 K and 37 K, respectively. In this paper, indexing of reflection is based on the orthorhombic lattice above $T_{\rm c}$.
As previously reported~\cite{sawa}, only $b$-axis in the orthorhombic phase monoclinically moves while both $a$- and $b$-axes retain their directions. Therefore, one can see the peak splitting associated with such an orthorhombic-to-monoclinic phase transition.
The monoclinic peak splitting resulting from two domains (hereafter called domain 1 and domain 2) starts at $T_{\rm c}$ with the evolution of the splitting with decreasing temperature. The splitting angle is almost saturated to $\Delta \omega = 0.07^{\circ}$ at the lowest temperature which is consistent with the previous reports~\cite{sawa,ninomiya}. The ratio of integrated intensities between domains 1 and 2 directly gave a volume ratio of the two domains as about 55:45 in the present experiment. 

\begin{figure}[t]
\begin{center}
\includegraphics[width=0.95\linewidth]{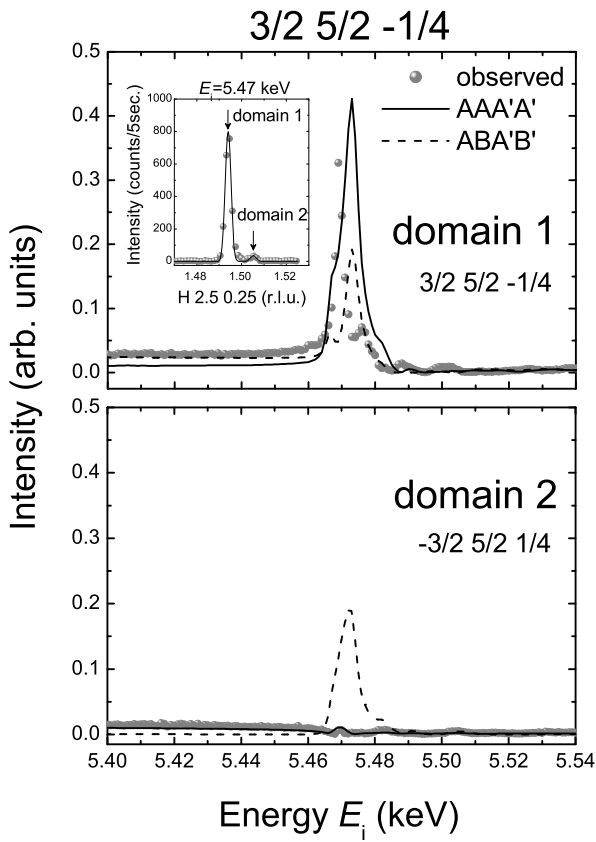}
\end{center}
\caption{Observed energy spectra of superlattice reflections $\frac{3}{2}\frac{5}{2}\overline{\frac{1}{4}}$, coming from domain 1 (upper) and domain 2 (lower). The dotted data show the observed energy spectra while solid and broken curves represent the calculated intensities based on AAA'A' and ABA'B' sequences, respectively. An inset in the upper panel shows the monoclinic splitting of the superlattice reflections which made the present RXS possible to unambiguously determine the charge-order pattern in the LT phase.}
\label{fig:3}
\end{figure}
Superlattice reflections also showed such a monoclinic splitting so that their energy spectra were measured for each of two domains as a function of incident x-ray energy ($E_{\rm i}$) across the K-absorption edge. The constraint of the present diffraction geometry allowed us to access to the following nine superlattice reflections,
$\frac{1}{2}\frac{1}{2}\overline{\frac{1}{4}}$,
$\frac{1}{2}\frac{3}{2}\overline{\frac{1}{4}}$,
$\frac{1}{2}\frac{3}{2}\overline{\frac{3}{4}}$,
$\frac{1}{2}\frac{5}{2}\overline{\frac{1}{4}}$,
$\frac{1}{2}\frac{5}{2}\overline{\frac{3}{4}}$,
$\frac{3}{2}\frac{3}{2}\overline{\frac{1}{4}}$,
$\frac{3}{2}\frac{3}{2}\overline{\frac{3}{4}}$,
$\frac{3}{2}\frac{5}{2}\overline{\frac{1}{4}}$ and 
$\frac{3}{2}\frac{5}{2}\overline{\frac{3}{4}}$, 
where the energy spectra were measured for each of both domains 1 and 2 at 7 K. Figure~\ref{fig:3} displays such an energy spectrum of the $\frac{3}{2}\frac{5}{2}\overline{\frac{1}{4}}$ (domain 1) and the -$\frac{3}{2}\frac{5}{2}\frac{1}{4}$ (domain 2)~\cite{lattice}. Dots in the upper and lower panels represent the experimental data for domains 1 and 2, respectively. 
One can clearly see remarkably different energy spectra between two domains. 
The inset in the upper panel of Fig.~\ref{fig:3} shows the monoclinic splitting of this superlattice reflection measured at $E_{\rm i} = 5.47$ keV where the most distinct difference between two domains was observed due to the critically enhanced contrast between V$^{4+}$ and V$^{5+}$ near the absorption edge. 
A systematic measurement was also made on other eight superlattice reflections. 

  In order to compare these observed energy spectra with calculations, we carried out the model calculation for the AAA'A' and ABA'B' stacking sequences. The present structure factor calculation was based on the results of the structural analysis performed by Sawa $\it{et~al.}$~\cite{sawa} and Ninomiya~\cite{ninomiya}. We used the same anomalous scattering factor f' and f'' of V$^{4+}$ and V$^{5+}$ ions as well as absorption factor as those used by Nakao $\it{et~al.}$~\cite{nakao,nakao2}
  Only two parameters, i.e., a scaling factor and an extinction correction parameter, were adjustable in the calculation. The Lorentz, temperature, and absorption factors were also taken into account. 

  Solid and dotted curves in Fig.~\ref{fig:3} represent the calculated results based on the AAA'A' and ABA'B' sequences, respectively. One can notice that the AAA'A' model reproduces the present observation much better than ABA'B'. In particular, the AAA'A' model very well reproduces the critical enhancement of the intensity on domain 1 and the weak intensity on domain 2. We also calculated the energy spectra of each domain of the other eight superlattice reflections and three fundamental reflections. All calculated spectra agree with observation very well and systematically. This fact leads to the unambiguous conclusion that the charge-order at the low temperature phase of NaV$_{2}$O$_{5}$ is the AAA'A' sequences along the $c$-axis.

It is essentially important that such obtained sequence AAA'A' simply consists of only two states A and A'. There is a striking experimental fact recently found in NaV$_{2}$O$_{5}$ which may support the AAA'A' model. Ohwada {\it et~al.} discovered the ``devil's staircase"-type behavior of the phase diagram of NaV$_{2}$O$_{5}$ at high pressures and low temperatures~\cite{ohwada1,ohwada2}. The ``devil's staircase"-type behavior is originally derived from a well-known theoretical ANNNI (Axial Next Nearest Neighbor Ising) model~\cite{ANNNI} forming a simple cubic lattice with Ising spins at each corner. Spins laying on the (001) plane are arranged ferromagnetically while along the [001] direction the first nearest neighbor inter-layer interaction is ferromagnetic ($J_{1} \geq 0$) and the second nearest one is antiferromagnetic ($J_{2} \leq 0$). These competitive interactions cause a frustration. This model surprisingly produces various types of higher-order commensurate phases with various types of spin modulations along the [001] layer-stacking direction as functions of temperature ($T$) and the interaction ratio $-J_{2}/J_{1}=\kappa$. For example, the phases with stacking modulations $q$ = 0, 1/4, 1/5 and 1/6 have spin configurations along the [001] direction as $\uparrow\uparrow\uparrow\cdot\cdot\cdot$ (all up configuration), $\uparrow\uparrow\downarrow\downarrow\cdot\cdot\cdot$ (up-up-down-down), $\uparrow\uparrow\uparrow\downarrow\downarrow\cdot\cdot\cdot$ (up-up-up-down-down) and $\uparrow\uparrow\uparrow\downarrow\downarrow\downarrow\cdot\cdot\cdot$ (up-up-up-down-down-down), respectively. The temperature-pressure phase diagram previously observed in NaV$_{2}$O$_{5}$~\cite{ohwada1,ohwada2} very well resembles the $T$-$\kappa$ global phase diagram of the ANNNI model. It is quite reasonable to understand that the low-temperature structure with the AAA'A' sequence determined by the present RXS experiment corresponds to the phase of $q$ = 1/4 with $\uparrow\uparrow\downarrow\downarrow$ configuration of the ANNNI model. The previously observed other phases with wave vectors $q_{\rm c}$ = 1/5, 1/6 and 0 are also easily understood by the stacking sequences AAAA'A'$\cdot\cdot\cdot$, AAAA'A'A'$\cdot\cdot\cdot$ and AAAA$\cdot\cdot\cdot$, respectively.

  By applying the RXS method to a monoclinically-split domain of NaV$_{2}$O$_{5}$, we have succeeded in unambiguous determination of the charge-order pattern of its low-temperature phase below $T_{\rm c}$ = 35 K. Observed energy spectra of nine sets of superlattice reflections show an excellent agreement with the calculation based on AAA'A' model and rule out the ABA'B' model. Such an experimental fact that a combination of only two states of A and A' forms the low-temperature phase offers a clue to understand the devil's staircase type behavior found in the pressure-temperature phase diagram of NaV$_{2}$O$_{5}$ with an aid of the ANNNI model. Further interesting issue is a microscopic mechanism for the AAA'A' sequence stabilized by a delicate balance between the first neighbor ferro-interaction favoring the A-A configuration and the second neighbor antiferro-one favoring A-A' causing frustration.

  We would like to thank Drs. Y. Wakabayashi, Y. Noda, M. Nishi, H. Seo and H. Fukuyama for stimulating discussions. 
  This work was supported in part by a Grant-In-Aid for Scientific Research from MEXT (Proposal 14740221). 
  A part of this work was supported by the NOP Project under the support of the Special Coordination Funds for Promoting MEXT. 
  The SRX-ray experiments were performed at the Photon Factory with the approval of Advisory Committee (Proposal 2001G254).




\begin{thebibliography}{99}

 \bibitem{isobe1}
 M. Isobe and Y. Ueda, J. Phys. Soc. Jpn. {\bf 65}, 1178 (1996).

 \bibitem{new-xray1}
 H. Smolinski, C. Gros, W. Weber, U. Peuchert, G. Roth, M. Weiden,
 and C. Geibel, Phys. Rev. Lett. {\bf 80}, 5164 (1998).

 \bibitem{seo-fukuyama}
 H. Seo and H. Fukuyama, J. Phys. Soc. Jpn. {\bf 67}, 2602 (1998).

 \bibitem{tsuda}
 K. Tsuda, S. Amamiya, M. Tanaka, Y. Noda, M. Isobe and Y. Ueda,
 J. Phys. Soc. Jpn. {\bf 69}, 1939 (2000).

 \bibitem{ohama1}
 T. Ohama, H. Yasuoka, M. Isobe, and Y. Ueda,
 Phys. Rev. {\bf B59}, 3299 (1999).

 \bibitem{fujii1}
 Y. Fujii, H. Nakao, T. Yosihama, M. Nishi, K. Nakajima,
 K. Kakurai, M. Isobe, Y. Ueda, and H. Sawa, 
 J. Phys. Soc. Jpn. {\bf 66}, 326 (1997).

 \bibitem{yosihama}
 T. Yosihama, M. Nishi, K. Nakajima, K. Kakurai, Y. Fujii, M. Isobe, C. Kagami and Y. Ueda, 
 J. Phys. Soc. Jpn. {\bf 67}, 744 (1998).

 \bibitem{ludecke} J. Ludecke, A. Jobst, S. van Smaalen, E. Morre, C. Geibel and H-G. Krane, Phys. Rev. Lett. {\bf 82}, 3633 (1999).

 \bibitem{boer} J. L. de Boer, SA. Meetsma, J. Baas and T. T. M. Palstra, Phys. Rev. Lett. {\bf 84}, 3962 (2000).

 \bibitem{nakao}
 H. Nakao, K. Ohwada, N. Takesue, Y. Fujii,
 M. Isobe, Y. Ueda, M. v. Zimmermann, J. P. Hill,
 D. Gibbs, J. C. Woicik, I. Koyama and Y. Murakami,
 Phys. Rev. Lett. {\bf 85}, 4349 (2000). 

\bibitem{ohama2}
T. Ohama, A. Goto, T. Shimizu, E. Ninomiya, H. Sawa, M. Isobe, and Y. Ueda, 
J. Phys. Soc. Jpn. {\bf 69}, 2751 (2000).

 \bibitem{sawa}
 H. Sawa, E. Ninomiya, T. Ohama, H. Nakao, K. Ohwada, Y. Murakami, Y. Fujii, Y. Noda, M. Isobe and Y. Ueda, J. Phys. Soc. Jpn. {\bf 71}, 385 (2002). 

 \bibitem{ninomiya}
 E. Ninomiya, Doctoral thesis, (Chiba University, 2003).

 \bibitem{grenier}
 S. Grenier, A. Toader, J. E. Lorenzo, 
 Y. Joly, B. Grenier, S. Ravy, L. P. Regnault, 
 H. Renevier, J. Y. Henry, J. Jegoudez and A. Revcolevschi, 
 Phys. Rev. B {\bf 65}, 180101(R) (2002).

 \bibitem{isobe2}
 M. Isobe, C. Kagami, and Y. Ueda, J. Crystal Growth {\bf 181}, 314 (1997).

 \bibitem{diamond}
 We used a conventional diamond anvil with 600 $\mu m \phi$ culet as the sample holder for the micro-crystals.

 \bibitem{lattice}
 A set of monoclinic lattice for domain 1 is assigned as (\vec{a}$_{1}$, \vec{b}$_{1}$, \vec{c}$_{1}$, $\gamma_{1} \ne 90^{\circ}$) while that for domain2 as (\vec{a}$_{2}$, \vec{b}$_{2}$, \vec{c}$_{2}$, $\gamma_{2} \ne 90^{\circ}$). In the present paper, following relations are defined: \vec{a}$_{1}$ = -\vec{a}$_{2}$, \vec{c}$_{1}$ = -\vec{c}$_{2}$, $\gamma_{1} = \gamma_{2} \ne 90^{\circ}$.

\bibitem{nakao2}
Sequential comments and reply on ref~\cite{nakao}. 
J. Garcia and M. Benfatto, Phys. Rev. Lett. {\bf 87}, 189701 (2001); J. E. Lorenzo, S. Bos, S. Grenier, H. Renevier and S. Ravy, Phys. Rev. Lett. {\bf 87}, 189702 (2001); H. Nakao, K. Ohwada, N. Takesue, Y. Fujii, M. Isobe, Y. Ueda, M. v. Zimmermann, J. P. Hill, D. Gibbs, J. C. Woicik, I. Koyama, and Y. Murakami, Phys. Rev. Lett. {\bf 87}, 189703 (2001). 

 \bibitem{ohwada1}
 K. Ohwada, Y. Fujii, N. Takesue, M. Isobe, Y. Ueda, H. Nakao, Y. Wakabayashi, Y. Murakami, K. Ito, Y. Amemiya, H. Fujihisa, K. Aoki, T. Shobu, Y. Noda and N. Ikeda, Phys. Rev. Lett. {\bf 87}, 086402 (2001).

 \bibitem{ohwada2}
 K. Ohwada, H. Nakao, H. Nakatogawa, N. Takesue, Y. Fujii,
 M. Isobe, Y. Ueda, Y. Wakabayashi and Y. Murakami,
 J. Phys. Soc. Jpn. {\bf 69}, 639 (2000).

 \bibitem{ANNNI}
 Per Bak and J. von Boehm, Phys. Rev. {\bf B21}, 5297 (1980), Per Bak, Rep. Prog. Phys., {\bf 45}, 587 (1982).

\end{thebibliography}
\end{document}